\documentstyle[12pt]{article}
\begin{document}

\setlength{\baselineskip}{0.30in}
\def\psl{p \hspace*{-0.5em}/}
\def\ksl{k \hspace*{-0.5em}/}
\newcommand{\Ga}{\Gamma}
\newcommand{\la}{\lambda}
\newcommand{\be}{\begin{equation}}
\newcommand{\ee}{\end{equation}}
\newcommand{\al}{\alpha}

\begin{flushright}
MPI-Ph/96-88\\
UM-TH-96-14\\
August, 1996\\
hep-ph/9609368
\end{flushright}

\begin{center}
\vglue .06in
{\Large \bf {Power Corrections and KLN Cancellations.}}\\[.5in]

{\bf R. Akhoury$^{(1,2)}$, L. Stodolsky$^{(2)}$, and V.I. Zakharov$^{(1,2)}$}\\
[.05in]

{\it{$^{(1)}$The Randall Laboratory of Physics\\
University of Michigan\\
Ann Arbor, MI 48109-1120}}\\[.15in]
{\it{$^{(2)}$Max-Planck-Institut fuer Physik\\
Fohringer Ring 6, 80805 Muenchen, Germany}}\\[.15in]

\end{center}
\begin{abstract}
\begin{quotation}
We consider perturbative expansions in theories with an infrared 
cutoff $\la$. The infrared sensitive pieces are defined as terms
nonanalytic in the infinitesimal $\la^2$ and powers of this 
cutoff characterize the strength of these infrared contributions.
It is argued that the sum over the initial and final
degenerate ( as $\la \rightarrow 0$) states  which is required by
the Kinoshita - Lee - Nauenberg theorem eliminates terms of order
$\la^0$ and $\la^1$. However, 
the quadratic and higher order terms in general
do not cancel. This is investigated in 
simple examples of KLN cancellations, of relevance to the
inclusive decay rate of a heavy particle, at the one loop level. 
\end{quotation}
\end{abstract}

\newpage

\section{Introduction }

Recently, there has been an upsurge of interest in the 
theoretical and phenomenological aspects of
power-suppressed infrared contributions \cite{renorm, bbz, vbsu, az}. 
In case of QCD, one analyzes
terms of order $(\Lambda_{QCD}/M)^k$ where $k$ ia an interger, $k\ge
0$, $\Lambda_{QCD}$ is the infrared
parameter of QCD, and $M$ is a large mass scale which could be say,
the total energy in $e^+e^-$ annihilation or in the
 decays of heavy particles it is the mass of a heavy
decaying quark.  The case of $k=0$ corresponds to logarithmic
divergences while positive $k$ corresponds to what we call
power-suppressed infrared contributions. Powers of $\Lambda_{QCD}$ can
emerge only through the so called dimensional transmutation.
Perturbatively, it is more convenient to
introduce an infrared cutoff parameter $\la$ which could for example be
an infinitesimal boson mass. Then the infrared sensitivity is 
studied by looking for the terms nonanalytical in $\la^2$ in the 
perturbative expansion of the physical observable under consideration.  
This is because while terms like $\la^2$ can come from
propagators, those nonanalytic in $\la^2$ can arise only as contributions 
from soft and collinear particles, thus signalling the importance
of such contributions.

The Kinoshita-Lee-Nauenberg (KLN)
 theorem \cite{k,ln} is a fundamental theorem
concerning  infrared singularities in quantum mechanics and quantum field
theory. The KLN theorem concerns infrared (IR) singularities
which may arise in perturbative expansions, that is terms like $\al ln\la$.
The theorem states that these terms disappear order by order in
perturbation theory once summation over
all initial and final states degenerate in 
 the limit $\la\rightarrow 0$ is performed.
A natural question is whether this cancellation extends to the
power-like suppressed IR terms of order $\la$, $\la^2 ln\la$, etc.
 In other words, whether the procedure
of summing over the degenerate states brings
 about not only finiteness but
 analiticity as well. Although the question appears to be of fundamental
importance in understanding the IR behaviour in quantum field theory, to
our knowledge, it has never been addressed in a systematic way.
Very recently \cite{az} it was conjectured that the KLN cancellation does
extend to power-like terms. There, in particular, an attempt was made
to use the theorem to understand the general reason behind the
cancellation of the terms of order $\la/M$ in inclusive physical
crossections.

In this paper
we will explore the  KLN cancellations at the level of IR power-like
corrections. In section 2 we present arguments, based on the
approach of Ref.\cite{ln} that a summation over
degenerate initial and final states, in general,
eliminates IR terms of order $\la^0$ (that is, $log(\la))$ and $\la^1$.
However, higher order nonanalytic terms like $\la^2 ln \la$ or $\la^3$ 
 are in general, not cancelled. 
In the rest of the paper
 we consider some examples which verify this conclusion
and elucidate the cancellation mechanism.
Our examples have relevance to heavy particle
inclusive weak decays in the one-loop appproximation
 and we study the infrared behaviour by giving an
infinitesimal mass, $\la$, to a boson. 
Then infrared sensitive pieces are identified
as terms non-analytical in $\la^2$ \cite{bbz}. In section 3
some simplifications relevant to the calculation of the 
nonanalytical terms in the correction to the mass and, of the
bremsstrahlung diagrams  are discussed. We also comment on the
connection between the description of power suppressed infrared
contributions in terms of the operator product expansion and that in 
terms of the nonanalytic terms obtained via Feynman graphs.
In section 4 we discuss the KLN cancellations in an elementary
example. Of particular 
technical interest is the implementation of the procedure of
averaging over the initial states directly in terms of Feynman graphs
( in contrast to the  original techniques which make
heavy use of disconnected graphs in the context of
old fashioned perturbation theory). Drawing from the  approach 
of Ref \cite{az} we illustrate this in the context of 
the example. Our findings imply that the procedure of
averaging over the initial states cannot be defined in a straightforward way
beyond linear in $\la$ terms. We conclude  with a brief summary.

\section{ Infrared Cancellations Following From the KLN Theorem }

In this section
we will outline an argument as to why in the KLN sum
not only the terms singular in the infrared cutoff but those linearly
dependant on it are cancelled as well. Our discussion follows that of
Ref.\cite{ln}. 

Consider the time evolution matrix, $U(t,t_0)$ in terms of which
the $S$ matrix can be written as, $U(0, \infty)^{\dagger}U(0, -\infty)$.
The Lee -Nauenberg probabilities $P_{ab}$ for a transition
from $a \rightarrow b$ are given by:
\be
P_{ab}~=~\sum_{[a]}\sum_{[b]}|\langle b|S|a\rangle |^2 
\ee
Here the sum is over all degenerate initial and final states
$[a]$ and $[b]$ respectively.This can be expressed in the form:
\begin{eqnarray}
P_{ab} ~=~\sum_m\sum_n (T^{+}_{mn})^{*}T^{-}_{mn} \label{pab}\\
T^{-}_{mn}~=~\sum_{[a]}\langle m|U(0,-\infty)|a\rangle^{*}
\langle n|U(0,-\infty)|a\rangle.
\end{eqnarray}
A similar expression holds for $T^+_{mn}$ involving a sum over the final
degenerate states $[b]$.
The content of the KLN theorem is the statement that both $T^+_{mn}$ and
$T^-_{mn}$ are separately free of infrared divergences of the type
$ln \la$ and hence so is $P_{ab}$ ( $\la$ is an infrared cutoff). 
We will argue now that under certain conditions 
to be discussed,
 the absence of singularities of the type
$ln\la$ in $T^{\pm}_{mn}$ separately implies that $P_{ab}$ also does not
contain any non analytical terms proportional to $\la$.

To see this in the lowest order,
 consider the expansion of $U(0,\pm\infty)$ 
 in terms of $gH_1$ the interaction Hamiltonian. Then to order $g$,
\be
T^{-}_{mn}~=~\sum_{[a]}\left ( \delta_{ma}\delta_{na} + g
{{\delta_{ma}(1-\delta_{na})} \over {E_a-E_n-i\epsilon}}(H_1)_{na}^{*} +
g {{\delta_{na}(1-\delta_{ma})} \over {E_a-E_m-i\epsilon}}(H_1)_{ma} + ...
\right ), \label{dsum}
\ee
with a similar expression for $T_{mn}^{+}$. Infrared divergences arise
from the vanishing of the energy denominators inside the brackets in
(\ref{dsum}). For example if $|a\rangle$ represents the state of a 
quark, and $|n\rangle$ represents that of a quark and a gluon,
then the latter
becomes degenerate with the former in the limit of a soft gluon. 
In this case the energy denominator $E_a-E_n$ vanishes proportional to
$\omega$ the soft gluon energy, giving rise to the usual infrared divergence.
A similar situation happens for the collinear divergence.
Now, upon performing the degenerate sum in (\ref{dsum}), it is easily
seen \cite{ln} that there are no terms of order $1/{\omega}$ as 
$\omega \rightarrow 0$ in $T^{-}_{mn}$ and the same is true for
$T^{+}_{mn}$ as well. Thus there are no terms of order $1/{\omega^2}$
in the summand of $P_{ab}$ (\ref{pab}),
 and since the phase space sum is proportional to 
$\omega d\omega$, there are no terms that diverge like $ln\la$.
( The same is true of any possible collinear divergence.) Now the
point is that since the terms of order $1/{\omega}$ are separately 
cancelled in $T_{mn}^{\pm}$, then not only are there no terms of order
$1/{\omega^2}$ in  the summand of $P_{ab}$, but there are none of
order $1/{\omega}$ as well. In fact,in general, the expansion of 
$(T^{+}_{mn})^{*}T^{-}_{mn}$ for small $\omega$
 starts off with the constant term .
Thus non analytical terms proportional to $\la$ are absent from
the KLN probabilities. A similar argument relying on the 
proven finiteness of the $T_{mn}^{\pm}$ separately in the
KLN sum and power counting generalizes to
higher orders in $g$ as well. 
There is however a condition which must be 
satisfied and it makes its appearance only beyond the leading order
brehmstrahlung diagrams. 

To prove the finiteness of $T_{mn}^{-}$ 
to all orders \cite{ln} one must consider three cases:
(1) Both $m$ and $n$ are in the degenerate set,$[a]$, (2) $m$ lies 
outside this set and $n$ may be contained in it, (3) the roles of 
$m$ and $n$ are reversed. Of these, case (1) follows from the unitarity
of the $U$ matrices, and case (3) is related to (2) by hermiticity. For
case (2) one uses induction. Since $U=U(0,\pm\infty)$ 
diagonalize the 
total Hamiltonian, then with $\hat{H}$ this diagonal quantity, we
can write:
\be
[ U, \hat{H}]~=~(gH_1+\Delta)U \label{delta}
\ee
where,the diagonal
$\Delta = H_0-\hat{H}$; $H_0$ is the free particle Hamiltonian
with bare masses and $\hat{H}$ is the same with physical masses. Thus,
for example for a massive quark, the matrix elements of $\Delta$
give the mass shift $\Delta m$. Let us denote by $O_{\alpha}$ 
the $\alpha^{th}$
term in the power series expansion of $O$, then by considering the 
appropriate matrix elements of (\ref{delta}) a recursion relation
for $T_{\alpha~mn}^{-}$ may be established
expressing it in terms of $T_{\beta}^-$
and  $\Delta_{\beta}$ for $\beta < \alpha$:
\be
T_{\alpha~mn}^{-}~=~{1 \over {E_a-E_m}}\left (
\sum_p H_{1~mp}T_{\alpha-1~pn}^-~+\sum_{\beta = 1}^{\alpha - 1}
\Delta_{\beta~mm} T_{\alpha-\beta~mn}^{-} \right ).
\ee
Together with the lowest order discussion given above , this
establishes,  not only the absence of infrared divergent terms proportional
to $ln \la$ but those linear in $\la$ as well, provided that
$\Delta_{\beta}$ has this property for all $\beta \leq \alpha -1$.
This in turn implies that the renormalization procedure must be
such as not to introduce any infrared sensitivity.

We conclude this general discussion with the following remarks:
(1)In this general discussion we have assumed that a suitable 
infrared regulator $\la$ exists whose vanishing produces the 
degeneracy. Moreover, our discussion of infrared sensitivity
is restricted to those arising due to this degeneracy alone.
 For explicit calculations, in field theories 
including the case of abelian gauge theories the regulator $\la$  can
for example to be a non vanishing but small (gauge) boson mass.

(2) The  statement on the absence of terms proportional to $ln\la$ and 
 $\la$ from the KLN sum is valid as we have just argued,
 in general in any field theory. 
Additional symmetries may require the vanishing 
of further terms nonanalytic in $\la^2$ like $\la^2 ln \la$ and so on.
An example of this is provided by gauge theories. Gauge invariance 
requires that if in the small $\omega$ expansion of $T_{mn}^{\pm}$ 
there are no terms of order $1/\omega$ then there are no constant terms
( order $\omega^0$ ) as well and hence the leading term is proportional
to $\omega$. This in turn implies that the leading non vanishing term
in $P_{ab}$ due to soft gauge particles,
that is non analytic in the infrared cutoff is proportional to
$\la^4 ln \la$.
 
(3) Let us consider the contribution of the collinear divergences
in the lowest order expansion (\ref{dsum}). Suppose that 
$|a\rangle$ represents a massless quark state and $|n\rangle$ 
that of the quark and a gluon. In the limit that the gluon is
moving parallel to the quark the two states become degenerate.
Then the energy denominator $E_a-E_n$ which goes like $(1-cos\theta)$
becomes of order $ (\theta^2 + \theta^4)$. Now in most situations
of physical interest the matrix element $(H_1 )_{na}$ itself for
 small $\theta$ becomes proportional to order $(\theta + \theta^3)$
due to helicity conservation at the vertex.
 Thus $T_{mn}^-$ becomes of order :
\be
T_{mn}^-~\sim~\sum_{[a]}{1 \over \theta}( 1 + \theta^2)
\ee
and similarly for $T_{mn}^+$. Thus since the $1/\theta$ term in the
degenerate sum for both $T_{mn}^{\pm}$  separately  vanishes and since
the phase space sum is proportional to $sin \theta d\theta$, we see
that collinear divergences in particular are not relevant 
for the terms linear in $\la$.

(4) The previous comment illustrates a feature that can be present
even for the case of soft infrared sensitivity. Namely the 
power counting may be modified if there is kinematic 
suppression , as for instance, in the case of the anomalous
magnetic moment interaction introduced as an example in section 4.
Here the cancellation is of the term proportional to
$\la^3$. The universal feature of the KLN theorem is the
cancellation of the  pole terms ( like $1/\omega$ ) in the degenerate sum. 

(5) It is known that in the calculation of $P_{ab}$  one must consider
interference  between graphs, some of which contain disconnected 
parts. In fact, in the covariant formulation of field theories
an infinite number of such contributions may have to be considered in 
a given order \cite{az}. Fortunately, at least for soft particles, 
it is possible to perform 
a rearrangement of the perturbation series so that the 
contribution of the purely disconnected pieces factor out. 
However, the disconnected pieces do leave behind a trace :
Their effect is encoded
in the rule that essentially every boson propagator  in a 
graph is supplemented by its complex conjugate \cite{az}.
In addition of course, as dictated by the KLN sum, every
emitted line is supplemented by an absorbed line.
In this approach, we may forget about  
disconnected graphs and deal directly with Feynman graphs suitably 
modified in this manner. We apply this procedure 
to an example in section 4.  

\section{ Computation of Terms Non-analytical in a Meson Mass$^2$.}

In this section we develop considerations which are
of relevance to the inclusive weak decays of
a  heavy fermion of mass $m$, ($m\gg \la$) which will be our principle
example. Because the total energy available for the decay is
of primary concern we study the analyticity of the mass of the heavy
particle in $\la^2$. For this purpose,since we are interested in matters of
principle, we are free to choose a model
which technically is most transparent. Thus we will introduce
an interaction of the decaying fermion with
various bosonic fields which couple to this fermion 
alone and not to the decay products.
 Of course, this scheme does not include the case of an
 interaction with a conserved charge but this is exactly the feature
which makes the calculations most simple.
Since it is only the initial particle that has new interactions the
loop effects are encoded in the wave function and mass renormalizations of
the initial particle. In particular, the renormalization of the mass affects
directly the phase space factor in the decay probability
and our problem is the evaluation of non-analytic contributions to a heavy
particle mass.

We now derive some results involving the
mass shift and the bremsstrahlung diagrams emphasizing the technical 
aspects and the calculational simplifications. In particular, we 
find that there is a simple prescription  for finding the nonanalytic
pieces in $\la^2$. These results
are used in the following section to demonstrate the KLN cancellations
and in section 5 some physical applications are mentioned.

We begin with an interaction of the form
$G \bar{\psi} \Gamma \psi \varphi$, where $\varphi$
represents the "light" meson of mass $\la$ and $\psi$
the heavy fermion of mass $m$. Consider 
the well-known expression for the one-loop correction to the mass:
\be
\Delta m \cdot\bar{w}w~=~{-i\over 4\pi^3}{G^2\over 4\pi }
\bar{w}(p) I(\psl )w(p)
\ee
where $G$ is the coupling constant,
$w(p)$ is the corresponding
spinor (which in our case, describes the heavy particle
at rest) and , finally, $I(\psl )$ is:
\be
I(\psl )~=~\int d^4k{\Ga(\psl -\ksl +m)\Ga \over
{(p-k)^2-m^2+i\epsilon}}{1\over (k^2-\lambda^2+i\epsilon)}
\ee
Here $\la$ denotes the mass of the meson, 
$\Ga$ is the vertex and we will concentrate for the moment
on the case $\Ga =\gamma_5, \gamma_5^2=-1$.

As usual, we would integrate first over $k_0$. The meson propagator has
poles at:
\be
k_0~=~\pm({\bf k}^2+\la^2)^{1/2}\mp i\epsilon.
\ee
For the fermion propagator we expand in $1/m$ and keep only the
nonrelativistic approximation for the kinetic energy. Then the poles are
at:
\be
k_0~=~ {{\bf k}^2\over 2m}+i\epsilon,~~~
k_0~=~2m+{{\bf k}^2\over 2m}-i\epsilon.
\ee
Here the first pole corresponds to a static interaction if we neglect
the $1/m$ terms altogether while the second pole corresponds to
the production of a pair in the intermediate state.

Now, we close the
contour in the lower half plane. Then we have two poles, at
 $k_0=\omega=({\bf k}^2+\la^2)^{1/2}$ and
at $k_0=2m+...$. A crucial point is that non-analytical terms
can be associated only with the pole at $k_0=\omega$. Because,
if $k_0\approx 2m$ then one can safely expand in $\la^2/m^2$ and
no non-analytical terms can arise. (In fact, this is
not entirely obvious since we are
going to keep the relativistic corrections to
terms non-analytical in $\la$.) 
We can check this considerable simplification by comparing with 
known results.
In particular, for the correction to the mass in case of
pseudoscalar interaction one has a closed form answer \cite{sch}:
\be
\Delta m~=~{G_P^2\over 4\pi}{m\over 8\pi}\cdot
\left(log{\Lambda^2_{UV}\over m^2}-{1\over 2}+{\la^2\over m^2}
+2{\la^2\over m^2}(1-{\la^2\over m^2})log{\la\over m}-2({\la\over
m})^3\sqrt{1-{\la^2\over2m^2}}cos^{-1}{\la\over2m}\right). \label{schw}
\ee
Let us check the terms of order
$\sim \la^2log(\la/m)$ and $\sim (\la/m)^3$. 
We use:
\be
\int d^4k {1\over k^2-\la^2+i\epsilon}\bar{w}{\Ga(\psl-\ksl+m)\Ga\over
(p-k)^2-m^2+i\epsilon}w~=~(+2\pi i)\int{d^3{\bf k}\over 2\omega}
\bar{w}{\Ga(\psl-\ksl+m)\Ga\over (-2m\omega+\la^2)}w
\ee
where the equality is understood in the sense that we keep only one pole
corresponding to $k_0=\omega -i\epsilon$, as is explained
above. Furthermore:
\be
\bar{w}{\Ga(\psl-\ksl+m)\Ga\over
(p-k)^2-m^2}w~=~\bar{w}{1\over
2m(1-\la^2/2m\omega)}w
\ee
where we have accounted for $(\gamma_5)^2=-1$,
$\bar{w}\ksl w=\bar{w}w\cdot \omega$ (that is,
$\bar{w}\gamma_{\la}w\neq 0$ only if $\la=0$). Moreover,
as far as the non-analytical terms are concerned
\be
\int{d^3{\bf k}\over 2({\bf k}^2+\la^2)^{1/2}}~=~\pi\la^2log(\la )
\label{simple}.
\ee
Note that the log factor in the r.h.s. is in fact negative while
the integral in the l.h.s. is apparently positive. As usual, this is due to
the subtractions needed to extract the non-analytic term. Combining the
factors we get for the leading non-analytic term 
\be
\Delta m~\approx~{-iG_P^2\over 16\pi^4}(2\pi i){1\over
2m}\pi\la^2log(\la)~=~{G_P^2\over 16\pi^2}{\la^2log(\la)\over
m} \label{quadr}
\ee 
which checks with (\ref{schw}).

For the next non-analytical term, we may expand the denominator above:
\be
{1\over 1-\la^2/2m\omega}~\approx~1+{\la^2\over 2m\omega}.
\ee
Moreover, again keeping only the nonanalytic terms,
\be
\int {d^3{\bf k}\over \omega}{\la^2\over \omega}~=~
-2\pi^2(\la ^2)^{3/2}.
\ee
And in this way we reproduce the cubic term correctly:
\be
(\Delta m)_{cubic}~=~{G_P^2\over 16\pi^4}(2\pi){1\over
8m^2}(-2\pi^2)(\la^2)^{3/2}
~=~-{G_P^2\over 32\pi}{(\la^2)^{3/2}\over m^2}
\label{cubic}.
\ee
Thus we see that the non-analytic terms (at least
in the one-loop approximation we are considering) can be 
obtained in a very simple way, i.e., they
arise only from the pole in the $k_0$ plane corresponding to a real meson,
in the intermediate state with a standard phase space
factor. On the other hand, all the kinematics in the rest of the graph can be
treated in a fully relativistic manner.
 That the infrared divergent terms are
obtained by considering the massless boson to be on shell is well known,
and what we see here is that the same is true for the other non analytic
terms as well. In general, it is a consequence of the Landau equations 
\cite{landau} concerning the analytic structure of Feynman graphs
which tell us that nonanalyticity in a mass results from putting
the corresponding line on shell.

It should be emphasized that the above procedure 
for $\Delta m$ is different from
 just  keeping those  terms
which are enhanced because of the small energy denominators.
( i.e., keeping only the pole term $1/\omega$ ) This difference
is an important point so let us discuss it further and examine
the origin of the terms of order $\la^2 ln \la$ and those of order
$\la^3$ from the viewpoint of a non relativistic expansion .

The pseudoscalar interaction is ideal to illustrate the point because
in this case the relativistic effects are not mere
corrections but rather the leading terms. Indeed, in
the non-relativistic limit:
\be
\bar{w}\gamma_5w~\rightarrow \bar{u} {({\bf \sigma \cdot k})\over 2m}u
\label{equiv}
\ee
where $u$ is a non-relativistic spinor, ${\bf k}$ is the 3-momentum
carried by the meson.
Eq (\ref{equiv}) expresses the equivalance (on the mass shell) of the
pseudoscalar and pseudovector couplings.
Therefore, in terms of the old-fashioned perturbation 
theory in the pole approximation the correction to the mass is:
\be
\Delta m\cdot\bar{u }u~=~{G_P^2\over 4m^2}\int {d^3{\bf k}
\over 2\omega(2\pi)^3}\bar{u}({\bf \sigma\cdot k}){1\over -\omega}({\bf
\sigma\cdot k})u
\ee
where the essential factors are the vertices (see Eq. (\ref{equiv}), the
energy denominator and the phase space of free mesons.
As a result, we have
\be
\Delta m~=~-{G_P^2\over 4m^2}\int {d^3{\bf k}|{\bf k}|^2\over 2\omega^2
(2\pi)^3}~=~-{G_P^2\over 32\pi}{(\la^2)^{3/2}\over m^2}.
\ee
In other words, we reproduce the cubic term above (see Eq. (\ref{cubic})).
 which we see is dominated therefore by the pole.

On the other hand, the leading, or $\la^2 ln \la$ term
 is not reproduced by
old fashioned perturbation theory in the pole approximation. The reason is
that the leading $\la^2log(\la )$ dependance is not due to
the pole $1/\omega$ in the amplitude but 
rather is an entirely relativistic effect.
If we consider a heavy quark expansion then we should have 
first eliminated the lower components of the Dirac spinors
 and thus arrive at the effective term
\be
L_{eff}~=~{G^2\over 2m}\bar{\psi}(x)\psi(x){\varphi}^2(x)
\label{lef}
\ee
where $\psi$ and $\varphi $ are the fermion and meson fields
(see, e.g., Ref. \cite{sch}, p. 411).
The correction to the mass proportional to $\la^2log(\la )$ 
arises clearly as a (perturbative)
 mesonic condensate - like contribution :
\be
\langle\varphi^2\rangle_{pert}~=~\int {d^3{\bf k}\over 2\omega(2\pi)^3}~=~
{\pi\la^2log(\la)\over (2\pi)^3}
\ee
where we used the integral (\ref{simple}) above. Combining it with Eq.
(\ref{lef}) we come again to (\ref{quadr}). Thus we see that 
it is nescessary to go beyond the most naive non-relativistic limit
to reproduce the  $\la^2 ln \la$ term.

We consider now the bremsstrahlung process, which is closely related 
to the self energy. In fact, previous calculations have found a
Bloch Nordsieck \cite{bn} type cancellation
 in the total ( weak decay ) width
to order $\la$ \cite{vbsu,bbz}.
We can treat the bremsstrahlung graphs in a fully
relativistic way, a la Feynman, without restricting
ourselves to the $1/\omega$ terms in the amplitude. Then
our "bremstrahlung" in this scalar case would not factorize any
longer and the probablity of radiation of a soft
particle is no longer a product of an emission
factor and of a non-radiative process. Indeed, the
relativistic corrections are in fact not related
to the pole, as emphasized above.
The most convenient way that we have found to
include the bremsstrahlung contribution for problems of
relevance to the infrared sensitivity in inclusive 
weak decays  is as follows. We will
assume, ( as is true in the realistic case), that integrating over
the decay products gives $\bar{w}\psl^5 w$ ( for a review
see for example \cite{review}) where
$w$ is the spinor describing the heavy fermion but treated now as
a field operator. Then we could write identically
\be
\bar{w}\psl^5 w~=~\bar{w}(\psl-m+m)^5w~\approx~
5m^4\bar{w}(\psl -m)w
\ee
Moreover the factor $(\psl -m) $ would cancel the propagator and we would
reduce the graph describing bremsstrahlung to one describing $\Delta m$.
The corresponding correction
to the total width can be shown to be 
\be
\delta \Ga_{tot}~|_{brem}~=~5{\Delta
m\over m}\Ga^{0} \label{former}
\ee
where $\Delta m$ now does include the leading $\la^2log(\la )$ term,
i.e. it corresponds to the fully relativistic
evaluation of $\Delta m$, the same as above. Thus 
contrary to the case of linear corrections to the 
mass of a heavy quark in a gauge theory \cite{vbsu,bbz} where a 
cancellation takes place, in this case 
,the corrections from the mass shift and the bremsstrahlung add 
instead of cancelling.

The above results  have an interpretation 
in terms of a formalism that is
close in spirit to the Operator Product Expansion. In fact we now
 apply it in a fully relativistic way, i.e., to Dirac fermions.
We will see that the approach
matches very nicely with
 the relativistic calculations discussed previously.
Very briefly the idea is the following: Consider inclusive
heavy quark decay and as mentioned in the previous paragraph,
the result of integration over the decay products
 gives $\bar{w}\psl^5w$  where the spinor $w$ is treated
now as an operator. The equations of motion read,
\be
(\psl-m-G_P\gamma_5\varphi)w~=~0
\ee
where we have a pseudoscalar field not interacting with
 the decay products.
 We can conclude that the correction 
to the total rate due to this interaction is related
to the matrix element:
\be
\langle \bar{w}G_P\gamma_5\varphi w\rangle_{pert}
\ee
When evaluating this matrix element we note that the pseudoscalar
interaction also enters through the exponential in:
\be
\langle \bar{w}G_P\gamma_5\varphi w\rangle_{pert}~=~
\langle T \bar{w}G_P\gamma_5\varphi w \exp{(i\int H_{int}dt)}\rangle.
\ee 
In this way we get:
\be
\langle \bar{w}G_P\gamma_5\varphi w\rangle_{pert}~=~2\cdot\Delta m
\ee
where the factor of 2 is purely combinatorial.
Moreover, as is explained above, $\Delta m$ contains the following:
\be
\Delta m~\sim ~G_P^2\langle\varphi^2\rangle~\sim~G_P^2\la^2log(\la)
\label{interpr}.
\ee
Thus the noncancellation of the $\la^2 ln \la$ term can also be
understood from an OPE like approach.
In the language of the physical processes, the factor of 2 is due to the
doubling of the effect of the mass shift through the radiation of mesons.
The crucial point here is that the nonanalytic in $\la^2$
terms are entirely due to the pole corresponding to a "real" meson in the
intermediate state. That is why the doubling exhibited by the equation above
is exact as far as the non-analytical terms are concerned, even if
the relativistic corrections are included.

\section{KLN Cancellations }

In this section we first check that the KLN cancellation up to terms
of order $\la$ found in section 2 on general grounds, actually takes 
place for single particle propagation at one loop. This example
would be relevant to checking the KLN theorem for processes
involving an incoming particle, such as in weak decays. 

The KLN cancellations are better understood in the language of old-fashioned
perturbation theory since the KLN theorem is rooted in Quantum Mechanics.
As discussed in section 2, 
the KLN procedure deals with small energy denominators which enter the
perturbative evaluation of wave functions:
\be
\Psi_{pert}~\approx~\Psi_0+\sum_{n}{(\delta V)_{n0}\over E_0-E_n}\Psi_n
\ee
where $\Psi_n$ is the unperturbed wave function and $E_n$ are
unperturbed eigenvalues of the energy.
Relevant to heavy particle decays are renormalization of the wave
function and correction to the mass. 
Cancellation of the renormalization of the wave function 
by bremsstrahlung graphs is a very universal
phenomenon which is to be expected
 and we concentrate, instead, on $\Delta m$.
Let us start with the scalar interaction which has a non-vanishing
non-relativistic limit:
\be
\Delta m\cdot\bar{w}w ~=~{-i\over 4\pi^3}{G_S^2\over 4\pi}\int d^4k
{\bar{w}(\psl-\ksl+m)w\over (p-k)^2-m^2}{1\over k^2-\la^2}.
\ee
 To examine the analyticity in $\la^2$ we keep (see section 3)
 only the contribution of the pole associated with
a "real" meson in the intermediate state:
\be
\Delta m~=~{1\over 2\pi^2}{G^2_S\over4\pi}\int{2m-\omega\over
(-2m\omega+\la^2)}{d^3{\bf k}\over 2\omega (2\pi )^3}  \label{scalar}
\ee
Furthermore, let us expand (\ref{scalar}) in $1/m$:
\be
{2m+\omega\over (-2\omega+\la^2)}~\approx~
{1\over -\omega}\left(1-{\omega\over 2m}   +{\la^2\over
\omega}\right)~\approx~{1\over -\omega -{\bf k}^2/2m}.
\ee
We see that we have an energy denominator which corresponds
to the transition from a heavy particle at rest to state of
the same particle and a meson of 3-momentum
${\bf k}$ with the kinetic energy of the heavy
particle taken into account to first order in $1/m$.

To get a KLN cancellation we should add a process with the energy
denominator of the opposite sign.
The problem is how to interpret this.
In the language of the Feynman graphs a natural step
is to add a graph with a complex conjugated propagator of the meson.
In fact this rule for the initial state summation was derived in
\cite{az} through an analysis of the relevant graphs. Thus one adds
\be
{-i\over k^2-\la^2+i\epsilon}~\rightarrow~\left({-i\over
k^2-\la^2+i\epsilon} \right)^{*}
\ee
Where, the overall sign of the propagator is changed
and $+i\epsilon$ is replaced
by $-i\epsilon$. Because of this change the poles associated with the
meson propagator are now located at: 
\be
(k_0)_{pole}~=~\pm({\bf k}^2+\la^2)^{1/2}\pm i\epsilon
\ee
Closing the countour of integration over $k_0$ in the lower half plane we
get what we denote by $(\Delta m)_{KLN}$
\be
(\Delta m)_{KLN}~=~{1\over 2\pi^2}{G_S^2\over 4\pi}
\int{d^3{\bf k}\over 2\omega(2\pi)^3} {2m+\omega\over
2m\omega+\la^2}\label{dmkln}
\ee
In other words $(\Delta m)_{KLN}$ can be obtained from the standard $\Delta
m$ by changing $\omega\rightarrow -\omega$ (in the amplitude but not in the
phase space factor since we have also changed the overall sign of the
propagator).
A physical interpretation of (\ref{dmkln}) can be given in terms of a
"KLN vacuum" which is populated with light mesons according to the phase
space factor $d^3{\bf k}/2\omega(2\pi)^3$. Then $(\Delta m)_{KLN}$
represents a forward scattering amplitude of the 
heavy particle at rest, off the mesons.
We see that indeed $\Delta m$ and $(\Delta m)_{KLN}$ cancel each other as
far as the $1/\omega$ terms are concerned. This then gives a
cancellation including terms proportional to $\la$.
 
 This is not true,
however, for the relativistic corrections.
Indeed, it is straightforward to see that adding the amplitudes with
 $+\omega$ and $-\omega$
does not eliminate quadratic terms:
\be 
{2m+\omega\over
2m\omega+\la^2}+{2m-\omega\over-2m\omega+\la^2}~=~{-{\bf k}^2\over
m(\omega^2+\la^4/4m^2)}.
\ee
This implies,
that because of the relativistic corrections
\be
\Delta m+(\Delta m)_{KLN}~\sim~G_S^2\la^2log(\la).
\ee

The above is in accordance with our discussion in section 2 in that
while order $\la$ terms have cancelled, $\la^2 ln \la$ terms have not.
We could attempt to improve the cancellation by looking
for an exact counterpart of $\Delta m$, e.g. some further averaging.
In an obvious manner, the 
energy denominator with opposite sign arises 
if one considers the collision
of a heavy particle and a light meson
 with opposite 3-momentum ${\bf k}$.
Thus, one may say that this state is indicated
by the small energy denominator itself.
Moreover, to ensure
a coherent addition of the amplitudes
 one has to introduce then a coherent
mixture of heavy particles in the initial state. Since we have already
adopted the "KLN vacuum" with light particles this might seem a
reasonable extention of the procedure. However,
as far as heavy decays are concerned, a moving particle lives
longer because of the time dilatation. Thus, at least at first sight we
should amend for this factor which would then again destroy the complete
cancellation of the two process (that is, with the meson in the initial
state and the standard mass correction). In this way we see that, if at
 all, one may continue with the KLN cancellations with relativistic
corrections included, only at the
 price of introducing artificial appearing procedures.

In any case, the KLN cancellation cannot be imposed on terms $\sim
\la^2log(\la )$. Indeed, let us go back to the example of $\Delta m$ due to
the pseudoscalar interaction considered in the previous subsection.
If we add $(\Delta m)_{KLN}$, that is the same expression as for $\Delta m$
but with $\omega\rightarrow -\omega$, the 
terms proportional to $\la^2log(\la )$ are
doubled, not cancelled:
\be
\Delta m+(\Delta m)_{KLN}
~\approx~{G_P^2\over 32 \pi^2}4{\la^2\over m}log(\la)~\approx~
{G_P^2\over 2m}2\cdot\langle\varphi^2\rangle_{pert} \label{cond}
\ee
Thus, in a sense the, first KLN non-cancelling terms are of two types,
 i.e., a kinematical correction due to the $1/m$ terms and
that due to a  mesonic "condensate" which produces
the $\la^2log(\la )$ terms.
It is worth noting that actually the difference between these two kinds of
terms might depend on the specific
 procedure of introducing the infrared sensitive parameter.
 In particular, the KLN summation was reduced first to adding
graphs with complex conjugated propagators in Ref \cite{az}. In this
reference the meson (photon in that case) is kept strictly massless and
IR parameters are introduced through limits of integration. In that case
adding amplitudes with $\pm\omega$ eliminates poles in any case and
the remaining terms look similar to local condensates.

It should be remarked here that when we
classify IR terms by power counting we
assume that there is no kinematical suppression.
The counting of powers of $\la$ may change when this
 further suppresion is properly included .Introduce, for
example an anomalous magnetic moment of the decaying fermion $\kappa$.
Then the corresponding leading contribution to $\Delta m$ is:
\be
\Delta m~=~-\al_{el}{\kappa^2\over 2m^2}(\la^2)^{3/2}.
\ee
This contribution arises from $1/\omega$ terms in the expression for
$\Delta m$ and is in fact subject to KLN cancellations.

In summary, the results of this section are in agreement with the 
general considerations of section 2 in the sense that that at least
the nonanalytical terms including those linear in $\la$ cancel
from the KLN sum. For purely kinematical reasons
it is difficult to define the KLN cancellation for terms of order $\la^2$.
Moreover, the non-analytical terms of order $\la^2log(\la )$ are not
subject to the KLN cancellation  in any case. Indeed, in 
a sense, these may be considered as arising, due to
near degeneracy of the vacuum manifested in a non-vanishing perturbative
condensate$\langle \varphi^2\rangle\sim \la^2log(\la )$. For QED
we found in section 2 that the first nonanalytic terms in the KLN sum
were not $\la^2 ln \la$ but actually $\la^4 ln \la$. 
Since each term with a given power of 
$\la$ is separately gauge invariant, this is consistent
with the idea that the first nonanalytic terms are of the condensate 
type (\ref{cond}), because in QED the gauge invariant condensate term
must be $\langle F_{\mu\nu}^2 \rangle$ which involves four powers 
of the momenta.

\section{Summary}

In summary, both by general arguments and by explicit examples
we have shown that KLN cancellations can be only partly extended to the
case of power suppressed non-analytical terms. Namely, what is cancelled are
the pole contributions. Generally speaking, they are responsible for the two
leading IR terms, that is, terms of order $\la^0,\la^1$.(Although the
counting can be modified by an explicit
suppression in the infrared region as in the
example of the anomalous magnetic moment). It is worth
pointing out here certain characteristics of the nonanalytic terms
contributing to $\Delta m$.
As discussed in section 3, $\Delta m$
among other nonanalytic pieces contains one proportional to
$\la^2 ln \la$ and another proportional to $\la^3$. The latter
term was shown to arise from the pole approximation in distinction
to the quadratic term which was relativistic in origin. The discussion
at the end of section 3 was in the context of a Bloch Nordsieck type
mechanism  where the entire $\Delta m$ contributions 
 which are of order $\la^2 ln \la$ and $\la^3$ are uncancelled.
However because of the origin of the $\la^3$ term and the fact that 
the KLN sum cancells terms with $1/\omega$ denominators, (see 
section 2), we see that it will be cancelled if both initial and 
final degenerate states are averaged. The same is not true of the 
term proportional to $\la^2 ln \la$ since its origin is the scalar
condensate. In this sense the cubic term is the analogue in the scalar
case of the anomalous magnetic moment interaction.

 At the technical level, in this paper
we also found a simple way to extract non-analytical terms from
Feynman graphs, and in addition given a straightforward way of
averaging over initial states in the KLN sum,
 utilizing and extending the results of Ref \cite{az}.
Although for the explicit examples ,
we concentrated on  weak decays of
 heavy particles it is rather obvious
that the results are of more general validity and apply to any process.
Also, two-loop and higher loop graphs can be
treated in a similar way. However, the
introduction of  a non-vanishing boson mass
restricts the applicability of our methods 
for explicit calculations in higher loops
to abelian theories only, for reasons of consistency.

\section{Acknowledgements} 
We would like to thank M. Beneke, V. Braun, V. Khoze, G. Marchesini, 
A. Mueller, A. Sirlin, G. Sterman, and A. Vainshtein for discussions.
This work was supported in part by the US Department of Energy.

\end{document}